\documentclass[prb,superscriptaddress,showpacs,twocolumn,amsmath,amssymb,nofootinbib,eqsecnum,tighten]{revtex4}

\usepackage{fancybox}
\usepackage{graphicx}
\usepackage{dcolumn} 
\usepackage{bm}      

\newcommand{\ket}[1]{\mbox{$|#1\rangle$}}

\newcounter{ichi}
\newcounter{ni}
\newcounter{sann}
\newcounter{yonn}
\newcounter{go}
\def\paragraph#1{} 

\begin{document}

\title{
Levitation and percolation in quantum Hall systems with
correlated disorder
}
\author{
Hui Song
}
\affiliation{Department of Applied Physics, Univ. of Tokyo
7-3-1 Hongo, Bunkyo-ku Tokyo 113-8656, JAPAN}
\author{
Isao Maruyama
}
\affiliation{Department of Applied Physics, Univ. of Tokyo
7-3-1 Hongo, Bunkyo-ku Tokyo 113-8656, JAPAN}
\author{
Yasuhiro Hatsugai
}
\email[]{hatsugai@sakura.cc.tsukuba.ac.jp}
\affiliation{Department of Applied Physics, Univ. of Tokyo
7-3-1 Hongo, Bunkyo-ku Tokyo 113-8656, JAPAN}
\affiliation{Institute of Physics, Univ. of Tsukuba
1-1-1 Tennodai, Tukuba Ibaraki 305-8571, JAPAN}

\date{\today}

\begin{abstract}
We investigate the integer quantum Hall system in a two dimensional
lattice model with spatially correlated disorder
by using the efficient method to calculate the Chern number proposed by
Fukui \textit{et\ al}.
Distribution of  charge density indicates
that the extended states at the center of each Landau 
band have percolating current paths,
which are topologically equivalent
to the edge states that exist in a system with boundaries.
As increasing the strength of disorder, 
floating feature is observed in an averaged Hall conductance
as a function of filling factor.
Its relation to the observed experiments is also discussed.

\end{abstract}


\maketitle

\paragraph{Introduction}

Two dimensional electron systems with disorder in a perpendicular
magnetic field have attracted much attention both experimentally
and theoretically. According to the scaling theory~\cite{Abrahams79},
all single-electron states are localized at zero
temperatures (Anderson localization). However, in the strong
magnetic field, all states form Landau levels
and there exist delocalized states at the center of 
each Landau level, which are extended and form percolating
paths in a context of network model~\cite{Trugman83,Kramer05}.
The transition from the Hall liquid 
to the Anderson insulator in the limit of $B \to 0$ 
was discussed by Khmelnitskii and Laughlin in terms of 
the levitation scenario, i.e., floating theory~\cite{Khmelnitskii84,Laughlin84}, where Landau bands merge together as
decreasing the magnetic field and the extended states float up towards
higher energy through the Fermi level.
This scenario gives a reasonable 
explanation for many experiments
and the global phase diagram was 
constructed based on it~\cite{Kivelson92}.
However, there still
remain many controversial arguments about the microscopic picture of
the floating theory~\cite{Haldane97}. 

\paragraph{controversial arguments?}
In the quantum Hall system, the topological structure of  
the ground state wave function plays an essential role
for quantization of the Hall conductance.
Each quantum Hall state can be assigned a topological
quantum number as the Chern number~\cite{Thouless82}, 
which is the Hall conductance in unit of $e^2/h$.
Although there are several works~\cite{Aoki81, Yang96, Sheng97, Sheng98,  Hatsugai99} to calculate the 
Chern numbers and succeeded to explain experimentally
observed direct transition,
the random potential,
which is spatially uncorrelated,
is not compatible with the continuum limit~\cite{Koschny01, Sheng01}.
Koschny $\mathit{et\ al}$. proposed that
unless the random potential has long range correlation, 
the floating theory in a lattice system
cannot be consistent with that in a continuous system~\cite{Haldane97}.
By investigating the localization length using a recursive Green-function method~\cite{MacKinnon81},
they found the negative-Chern number cannot be created
if the correlation length of randomness is long enough. 
However, since the main part of their study is to calculate Green-function,
topological nature of extend states is unclear.

\paragraph{Goal of this paper}
To study topological nature of one-electron states, 
we numerically calculate both the spatial distribution of wave functions
and the Chern number in this paper.
Here, we make a spatially correlated random potential 
in order to explore features of a continuous system in a lattice system.
In previous studies on the floating scenario with the correlated
disorder~\cite{Koschny01, Sheng01, Pereira02, Koschny03, Koschny04},
the connection between the spatial probability density and the Chern number
has not been focused on.
To calculate the Chern number,
an efficient numerical method recently proposed by Fukui \textit{et\ al.}~\cite{Fukui05}
is employed.

\paragraph{Model and Method}
A tight-binding Hamiltonian on a square lattice with
nearest neighbor hopping
is given by 
\begin{equation}
H = t\sum_{\langle r,r' \rangle}c_r^{\dagger} e^{-i \theta_{r,r'}} c_{r'} + \mathrm{H.c.} 
  + \sum_{r}w_r c_r^{\dagger}c_r.
\end{equation}
The magnetic flux per plaquette $\phi$ is given by
$\sum_{plaquette}\theta_{r,r'}=2\pi \phi$, where the summation runs over
four links around a plaquette.  We choose the Landau gauge, that is,
$\theta_{r,r+\hat{x}} = 0,\ \theta_{r,r+\hat{y}}=2 \pi \phi r_x $.  
We set the units of energy to $t=1$, and mostly set
the magnetic field per one plaquette $\phi = 1/16$.  A
random potential, $w_r$, is spatially Gaussian correlated as
\begin{equation}
 \langle w_r w_{r'} \rangle = W~e^{-|r-r'|^2/\lambda^2},
\end{equation}
with mean value $\langle w_r \rangle = 0$,
where $\langle\ \rangle$ denotes the ensemble average
over different randomness realizations.
The distribution of the random potential is characterized by the 
correlation length $\lambda$ and the strength $W$. 
Although the $\lambda$ dependence is important\cite{Koschny03,
  Koschny04}, we will focus on the dependence on $W$ in the present paper
and $\lambda$ is fixed to a large value enough to consider
a continuous system in a lattice system.

In order to calculate Hall conductance in a system without the translational symmetry,
we impose a generalized boundary condition on every randomness 
realization, by requiring 
$T_\mu(L_\mu)\ket{\psi} = \exp (i\theta_\mu)\ket{\psi}\ (\mu=x,y)$, where 
$T_\mu(L_\mu)$ is the magnetic translation operator for a displacement $L_\mu$
and $\theta_\mu$ is the boundary-condition phase.
The system size is $L=L_x \times L_y$.
The Hall conductance $\sigma_{xy}$
can be calculated by 
$\boldsymbol{\theta}$ dependence of the wave function~\cite{Thouless82, Niu85},
where $\boldsymbol{\theta}$ is a vector defined as $\boldsymbol{\theta}=(\theta_x,\theta_y)$.
As for fixed particle density 
we define a ground state multiplet $\boldsymbol{\psi}(\boldsymbol{\theta})$,
which is a $L \times q$ matrix written as 
$\boldsymbol{\psi}(\boldsymbol{\theta})=(\ket{\psi_1(\boldsymbol{\theta})},....,\ket{\psi_q(\boldsymbol{\theta})})$
with $H(\boldsymbol{\theta}) \ket{\psi_j(\boldsymbol{\theta})}= \epsilon_j(\boldsymbol{\theta}) \ket{\psi_j}$~\cite{Hatsugai0405}.
Here, $q$ is a number of one-particle states below the Fermi energy.
Then we define a non-Abelian Berry connection $\mathcal{A}$ as
$\mathcal{A} 
= 
\boldsymbol{\psi}^{\dagger}d\boldsymbol{\psi}
$.
Finally,
the associated Chern number is given by
\begin{equation}
C_{\psi}=\frac{1}{2 \pi i}\int_{S} \mathrm{Tr}~d\mathcal{A},
\end{equation}
where $S$ is a two dimensional torus parameterized by
the boundary-condition phases $\theta_x$ and $\theta_y$. 
Recently, Fukui $\mathit{et \ al}$.~\cite{Fukui05} 
developed an efficient method to calculate a Chern number 
on a discrete momentum space by applying
the geometrical formulation of topological charges in the lattice gauge
theory. 
As an advantage of their method,
one can obtain the Chern number accurately
even if one uses coarsely discretized mesh points in the momentum space.

\begin{figure}[!htb]
\begin{center}
\includegraphics[width=8.0cm]{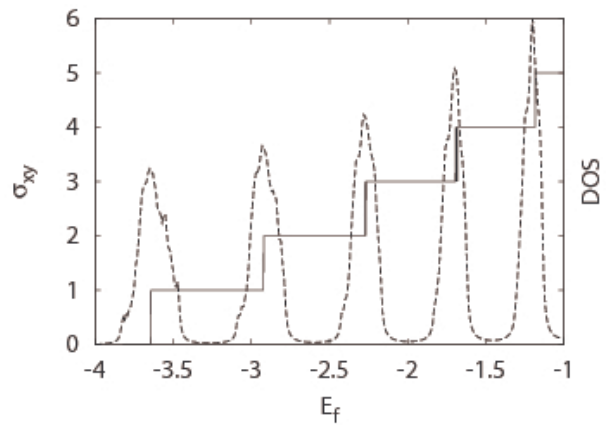}
\end{center}
\caption{ The Hall conductance $\sigma_{xy}$ (solid line) as a
  function of Fermi energy $E_f$, and density of states versus energy
  $E$ (dashed line), at a lattice size 32 $\times$ 32. The magnetic
  flux per plaquette ($\phi$) is $1/16$, the correlation length
  ($\lambda$) is 8, and disorder strength ($W$) is 0.01.}
\label{Ch-E}
\end{figure}

\begin{figure}[!bht]
\begin{tabular}{lr}
 \resizebox{40mm}{!}{\includegraphics{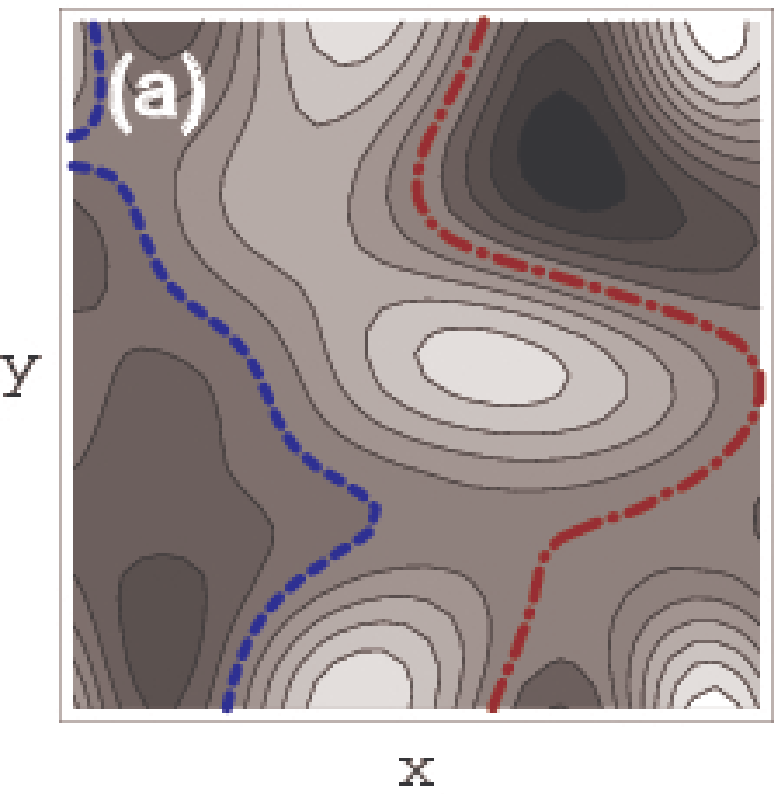}}
 \resizebox{40mm}{!}{\includegraphics{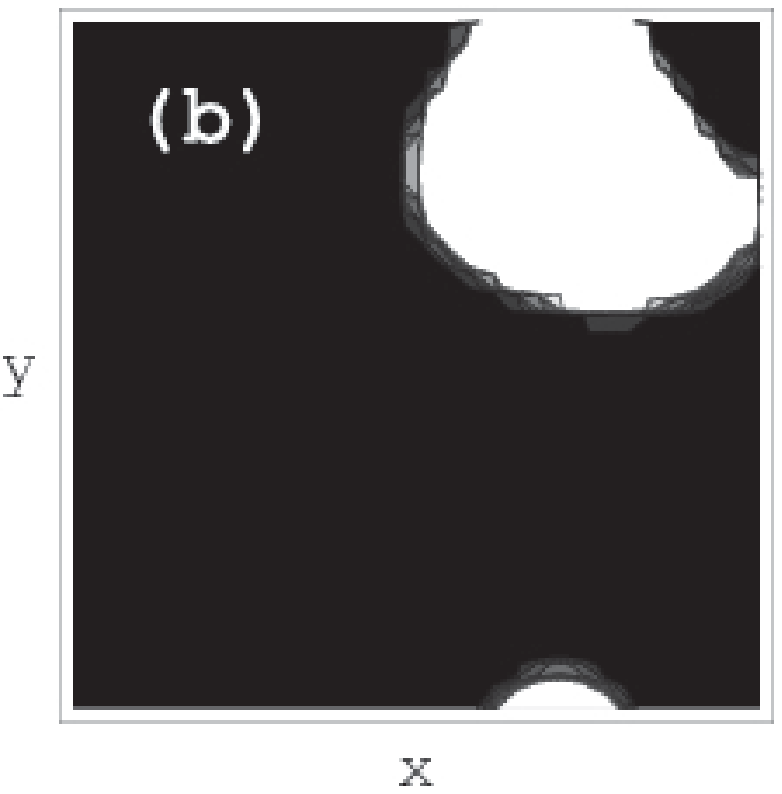}} \\
 \resizebox{40mm}{!}{\includegraphics{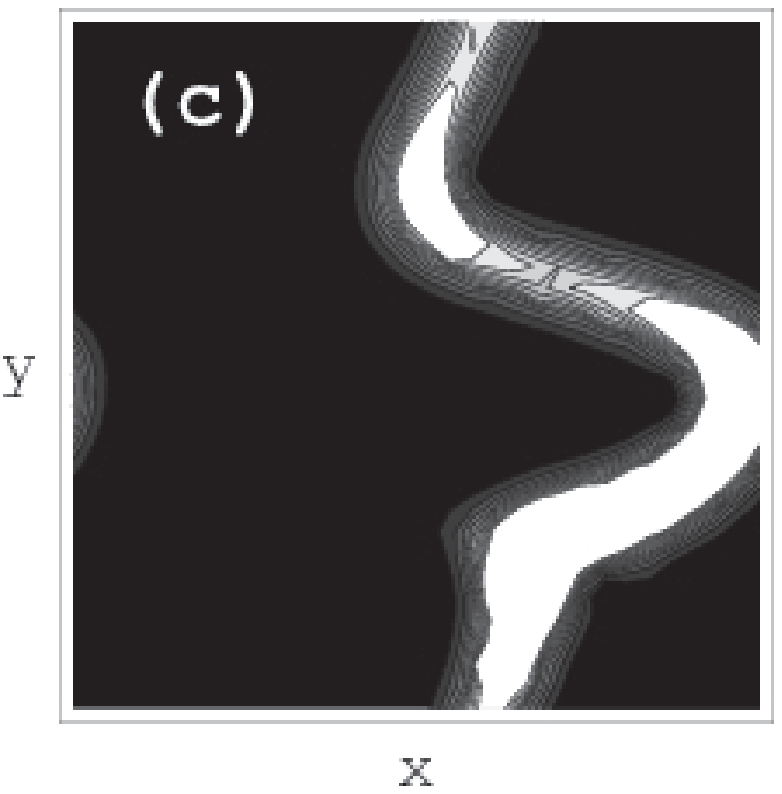}}
 \resizebox{40mm}{!}{\includegraphics{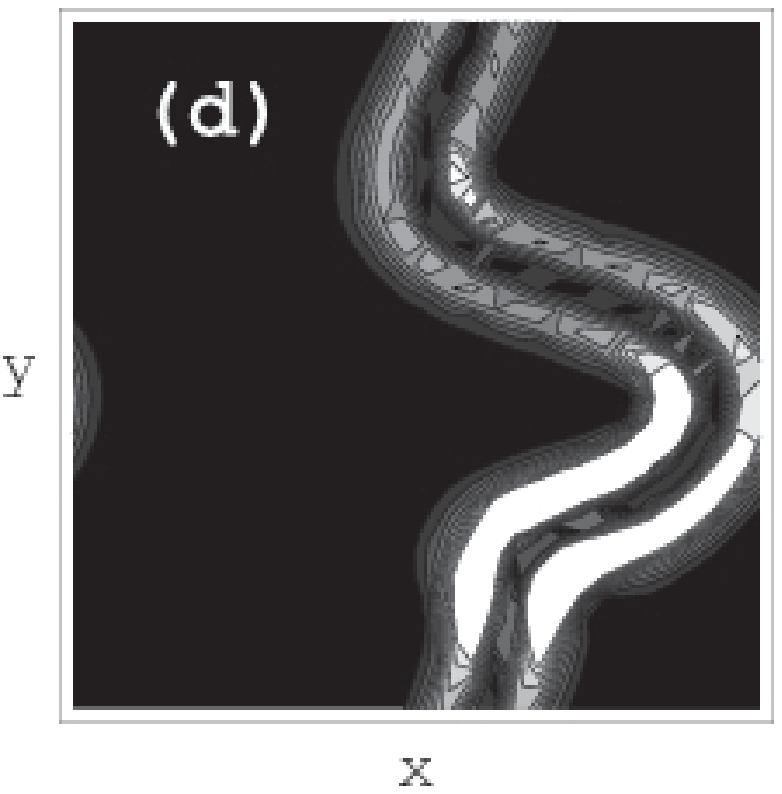}} \\
\end{tabular}
\caption{(Color online) Contour plot of randomness realization for $\lambda=8$ (a). 
The density probability distribution of several states, 
the localized state in the tail of lowest Landau band (b),
the critical state in the lowest Landau band (c),
and
the critical state in the second lowest Landau band (d).
The contour line indicated with the dashed-dotted line in (a) 
corresponds to the energy of the state of (c).
The dotted line in (a) is the contour line with the same energy.
}
\label{config}
\end{figure}

\paragraph{Topological meaning of extended states}
When randomness is introduced to a pure system, 
the Landau bands will broaden. 
At the tails of each band, all one-electron states become
localized due to the Anderson localization. 
On the other hand,
there still exist extended states 
at the center of each Landau band.
Figure~\ref{Ch-E} shows the dependence 
of Hall conductance $\sigma_{xy}$ and density of
states (DOS) on the Fermi energy $E_f$
for the given randomness realization 
that is plotted in Fig.~\ref{config}(a)
as a contour map. 
The strength of the randomness $W$ is set to 0.01.
Nevertheless the energy gap induced by the randomness 
is small,
numerical calculations of the Hall conductance are stable.
As shown in Fig.~\ref{Ch-E},
the Hall conductance jumps
whenever the Fermi energy goes across
the center of each Landau band.
The extended states in the center of each Landau band
can be classified into two classes; one possess non-zero 
Chern number, which is called a critical state, and the other does not.
There is only one critical state in the middle of the 
lowest band as well as the second lowest band. In general, there can exist several 
critical states at the center of Landau band.

To study topological structure of a wave function, 
the contour maps of the probability densities for several states
are shown in Fig.~\ref{config}(b), (c) and (d).
Note that we adopted the periodic boundary condition.
As shown in Fig.~\ref{config}(b), the probability density of the 
localized state in the lowest band tail, 
is concentrated around the valley 
of the random potential. 
On the other hand, 
the critical state as seen in Fig.~\ref{config}(c) or (d)
forms a percolating path along the $y$-direction,
which is associated with 
the percolation of equipotential lines~\cite{Trugman83, Kramer05}.
As shown in Fig.~\ref{config}(d),
two percolating paths and one nodal line between them are observed for  the second
Landau bands.
This distribution reminds us that 
the wave functions in the Landau gauge
can be described by the Hermite function $H_n(r)$,
which has $n-1$ nodal lines.
Since other extended states which do not have
nonzero Chern number 
show the similar spatial distribution,
the percolating path is thought to be a characteristic feature of extended states.

\begin{figure}[bht]
\begin{center}
\includegraphics[width=6.0cm, height =4.0cm]{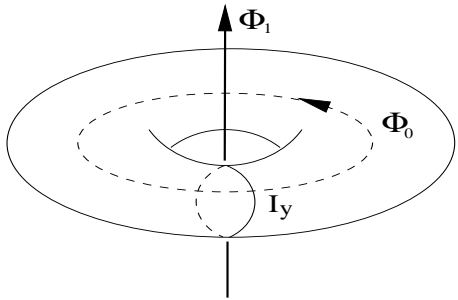}
\end{center}
\caption{
Two possible insertion of Aharonov-Bohm flux through
a torus.
}
\label{torus}
\end{figure}

\paragraph{percolating paths.}
There is a relationship between
the topological character and the spatial distribution of
wave function.
Let us consider adiabatic processes of inserting the 
Aharonov-Bohm (AB) flux into the holes of the torus 
(See Fig.~\ref{torus}).
Here it should be noted that the topology of 
the square lattice used in the present paper is a torus
due to the periodic boundary condition.
The penetration of flux $\Phi_0$ ($\Phi_1$) 
into the hole inside (outside) of the torus,
which is equivalent to twisting the boundary phases
$\theta_y$ ($\theta_x$),
could induce an infinitesimal electric field 
in the $x$($y$)-direction.
The charge-current response to the electric field
is given by the Byers-Yang formula:
\begin{eqnarray*}
 I_x = \frac{\partial E}{\partial \Phi_0},
\quad
 I_y = \frac{\partial E}{\partial \Phi_1},
\end{eqnarray*}
where $E$ is the total energy of the system.

\begin{figure}[!htb]
\begin{tabular}{lr}
  \includegraphics[width=8.0cm, height=5.0cm]{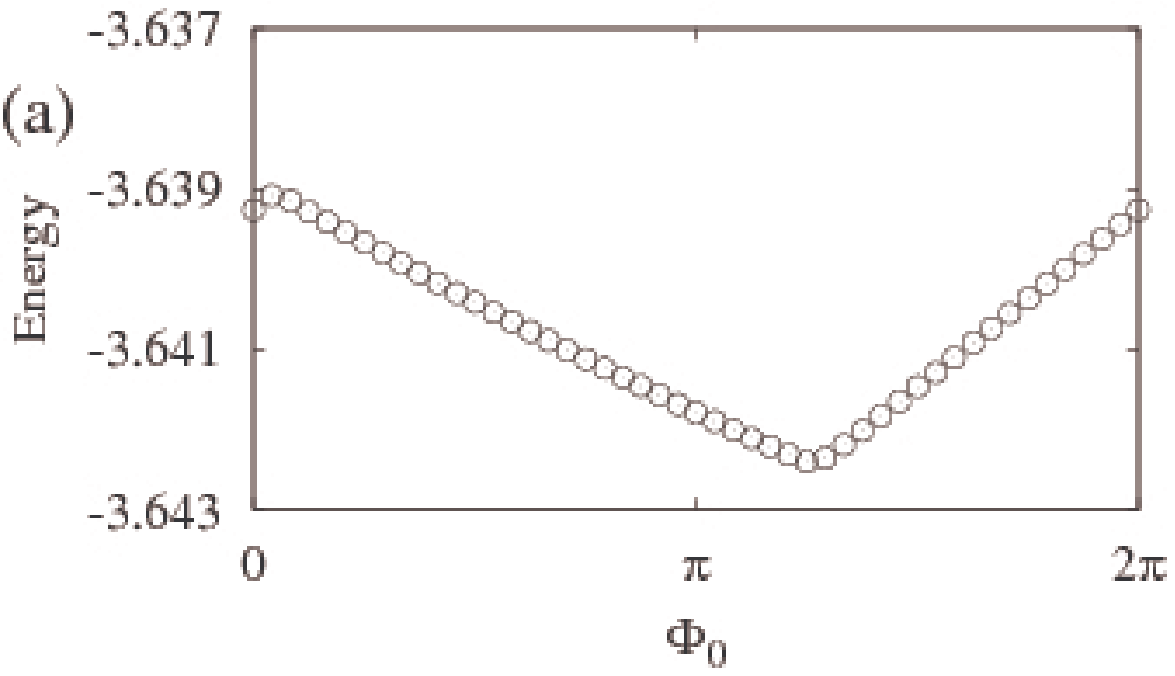}\\
  \resizebox{40mm}{!}{\includegraphics{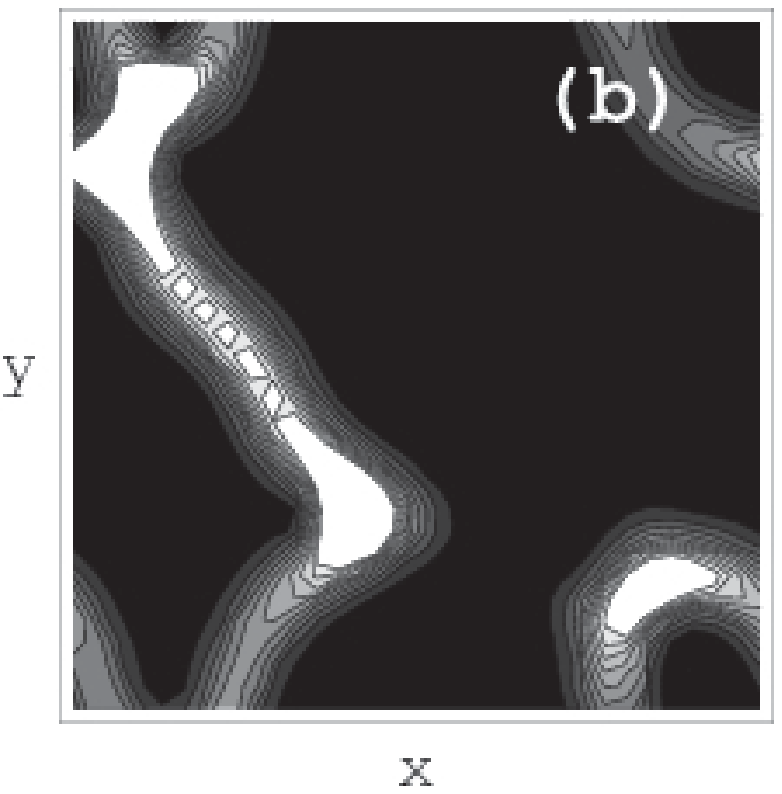}}
  \resizebox{40mm}{!}{\includegraphics{fig2c_4c.eps}} \\
\end{tabular}
\caption{(a) Energy dependence on AB flux.
The distribution of probability density of 
the critical state in the lowest band,
when $I_y > 0$ (b) and when $I_y > 0$ (c).
}
\label{E-Phi}
\end{figure}

When the AB flux in the unit of the magnetic flux quantum $h/e$
is adiabatically turned on from $0$ to $2\pi$,
the spatial distribution of the probability density changes drastically, 
depending on the sign of $I_y$
as shown in Fig.~\ref{E-Phi}.
Fig.~\ref{E-Phi}(a) shows the energy 
as a function of the AB flux $\Phi_0$
for the critical state in the lowest band.
Fig.~\ref{E-Phi}(b) or (c) shows the typical probability
density when the energy increases (decreases) monotonously, 
i.e., $I_y  > 0\ (I_y < 0)$.
It turns out that
this probability
density corresponds to
the percolating path along which the current flows.
Actually, 
Fig.~\ref{E-Phi}(b) or (c) corresponds to
the contour line shown as the dotted line (or the dashed-dotted line) 
in Fig.~\ref{config}(a).
These two contour lines, which corresponds to the percolating paths, 
indicate the same energy in the random potential.
While the $\Phi_0$ dependence is drastic,
we can hardly find a variation
with $\Phi_1$
in the total energy and 
the distribution of probability density.
This insensitivity to the flux $\Phi_1$ is
attributed to
the fact that there is no percolating
path in the $x$-direction. 
We also found the localized state 
is insensitive to 
both $\Phi_0$ and $\Phi_1$,
because there is no current path in the $x$ direction nor in the $y$ direction.
The correspondence between the percolation path and
the direction of the current can be
understood in the view of a classic picture.
In the system with boundaries,  
the electrons form skipping orbits
along the boundaries corresponding to the
cyclotron motion.
The boundaries in this picture are formed by 
 mountains of the random potential,
which correspond to the dotted and dashed-dotted lines
in Fig.~\ref{config}(a).
In this meaning, the percolation paths are thought to be 
the edge states.

\begin{figure}[!htb]
\begin{center}
\includegraphics[width=8.0cm]{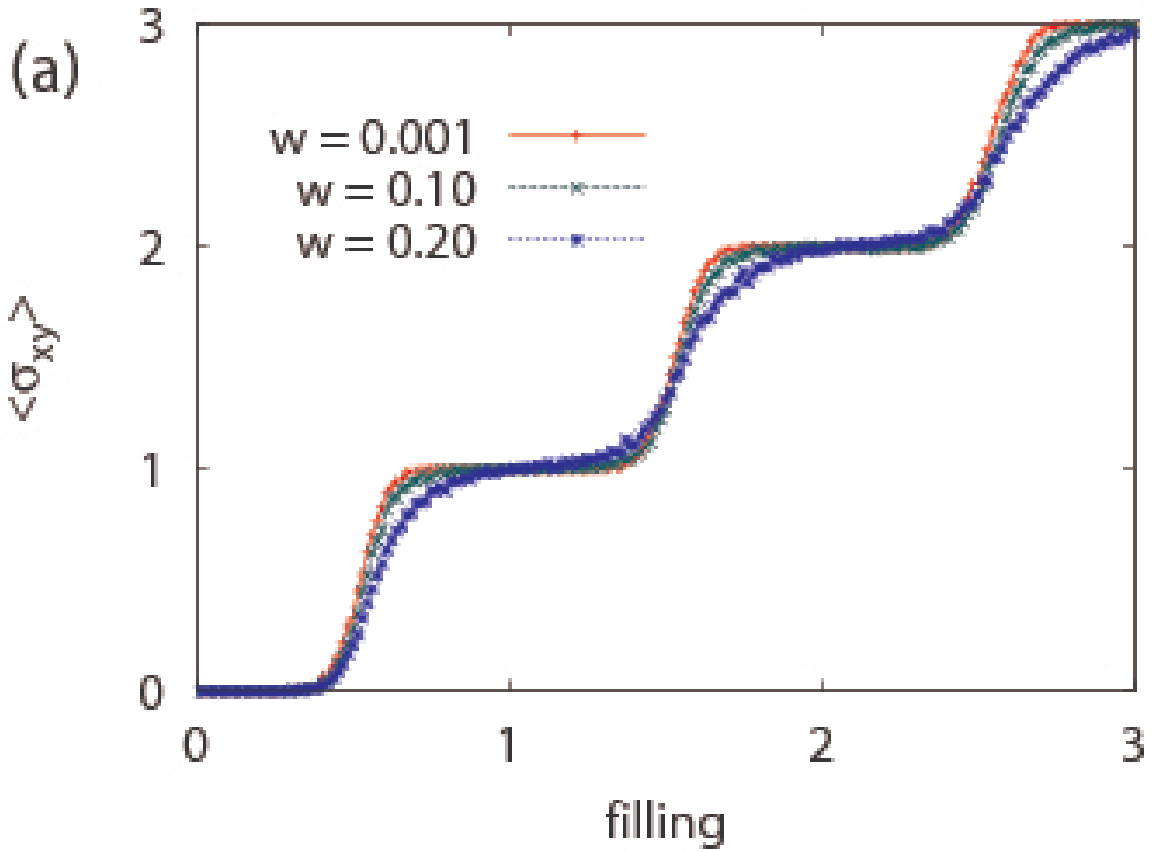}
\includegraphics[width=8.0cm]{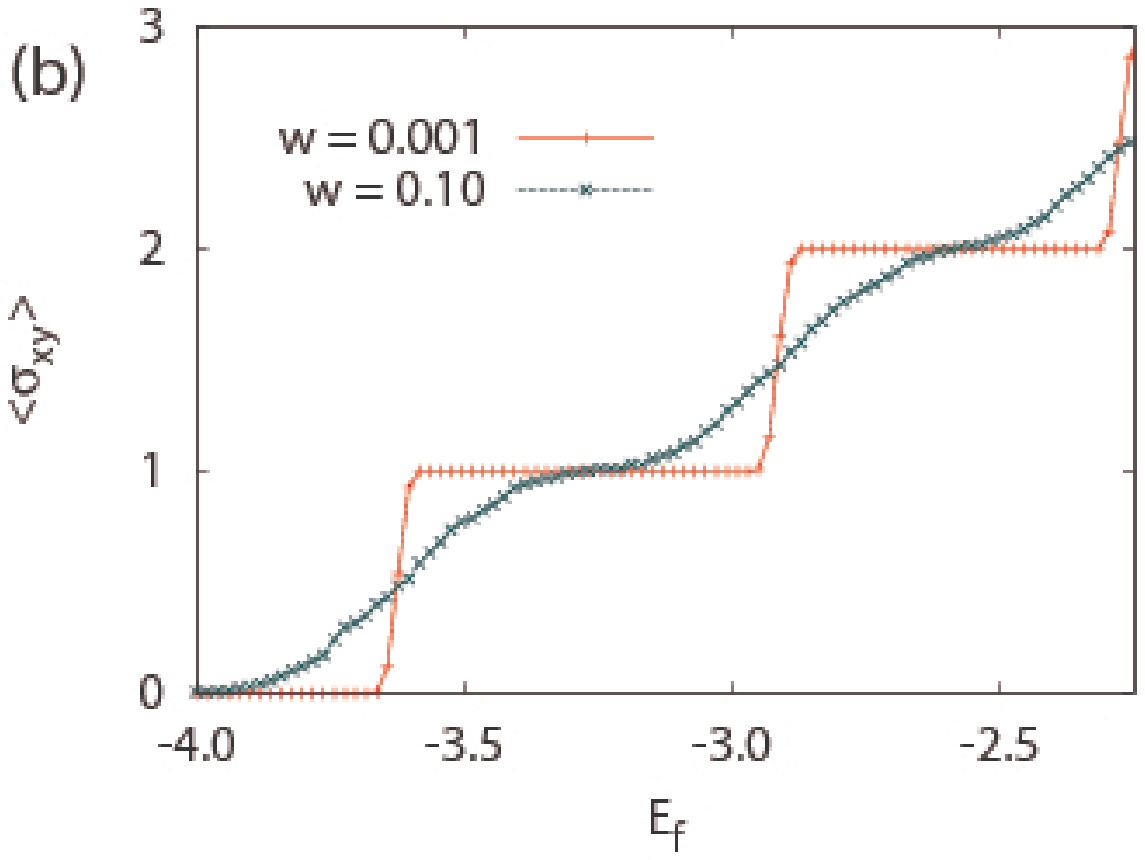}
\end{center}
\caption{(Color online) The averaged $\sigma_{xy}$ as a function of
  filling factor (a) and Fermi energy $E_f$, at various strength of
  randomness (b).  The correlation length ($\lambda$) is 8.  The assemble
  average is taken by 320 randomness realization.  }
\label{fig:Lev}
\end{figure}

\paragraph{fig5}
To take into account experimental situation, 
we consider averaged Hall conductance 
$\langle \sigma_{xy} \rangle$, which is
defined as an average over different randomness realizations.
Effect of increasing the disorder strength $W$
on $\langle \sigma_{xy} \rangle$
as a function of the filling factor
is shown in Fig.~\ref{fig:Lev}(a). 
With increasing $W$, one can see
the broadening effect on the steps of the Hall conductance.
Moreover, the step in the lowest Landau band
systematically shifts to the higher filling.
This behavior is consistent with the floating picture.
On the other hand, the behavior of  $\langle \sigma_{xy} \rangle$
as a function of energy is entirely different
as shown in Fig.~\ref{fig:Lev}(b).
Sharp steps at a weak randomness $W=0.001$ are broadened completely by randomness
in Fig.~\ref{fig:Lev}(b).
This is because 
the randomness is not only
broadening the Landau bands, 
but also shifts the position on Landau bands
depending 
on the mean value of random potential in each realization.
This suggests,  it is difficult to detect the 
float-up in energy when the sample measured in experiment has many 
domain walls.

\paragraph{Conclusion}
In conclusion, we have numerically calculated the Hall conductance,
i.e., the Chern number, on a 2D system with a uniform magnetic field and a
spatially correlated random potential.  
We have found that numerical method proposed by  Fukui \textit{et\ al}. is suitable to 
obtain the Chern number in a system with a random potential. 
We have also investigated the
spatial distribution of the wave functions to study the topology 
of extended states and localized states.  The
probability density of the extended states 
at the center of the Landau bands
forms a current path associated with the percolation of equipotential
lines.
On the other hand, 
for the localized states in band tails,
the charge density accumulates around 
the valley (or mountain) of the potential. 
The  topological difference of wave function results in the different response to the AB flux:
the energy of the extended state is sensitive 
to the AB flux, while the
AB flux has almost no effect on the localized state.
Finally, after taking an average of the Hall conductance over
different randomness realizations, floating 
feature to higher filling is observed.
\paragraph{future works}
\paragraph{Discussion about Dirac fermions}

\begin{acknowledgements}
This work was supported by Grant-in-Aid from
the Ministry of Education, No. 17540347 from JSPS, No.18043007 on
Priority Areas from MEXT and the Sumitomo Foundation. The computation has been partly
done using the facilities of the Supercomputer Center, 
University of Kyoto
and the facilities of the Supercomputer Center, 
Institute for Solid State Physics, University of Tokyo.
\end{acknowledgements}

\end{document}